\documentclass[12pt]{article}
\usepackage{geometry}
\usepackage{hyperref}
\usepackage{url}
\usepackage{latexsym}
\usepackage[centertags]{amsmath}
\usepackage{amssymb}
\usepackage{amsthm}
\usepackage{enumerate}
\usepackage{graphicx}
\usepackage{amsfonts,amssymb}
\usepackage{amsmath}

\usepackage[all]{xy}  
\usepackage{stmaryrd}

\usepackage{tikz}
\usepackage{tikz-cd}
\usetikzlibrary{shapes.misc,arrows,decorations.markings}
\usetikzlibrary{matrix}

\newtheorem{definition}{Definition}
\newtheorem{proposition}{Proposition}

\newtheorem{corollary}{Corollary}

\newcommand {\cM}{\mbox{${\mathcal M}$}}

\newcommand{\NN}{\mathbb{N}}
\newcommand{\R}{\mathbb{R}}
\newcommand{\LL}{\mathbb{L}}
\newcommand{\Cinf}{C^\infty }

\newcommand{\itSigma}{\mathit{\Sigma}}
\newcommand{\itDelta}{\mathit{\Delta}}
\newcommand{\Ric}{\mathrm{Ric}}
\newcommand{\ev}{\textrm{ev}}
\newcommand{\Hom}{\mathrm{Hom}}
\newcommand{\Der}{\mathrm{Der}}

\begin{document}

\title{Functorial differential spaces and the infinitesimal structure of space-time}

\author{ Leszek Pysiak \\ {\small Institute of Mathematics and Cryptology,} \\ {\small Military University of Technology, Kaliskiego 2, \mbox{00-908} Warsaw, Poland} \\[2ex]
		 Wies\l{}aw Sasin \\ {\small Faculty of Mathematics and Information Science,} \\ {\small Warsaw University of Technology, Koszykowa 75, \mbox{00-662} Warsaw, Poland} \\[2ex]		 
		 Michael Heller \\ {\small Copernicus Center for Interdisciplinary Studies,} \\ {\small Jagiellonian University, Szczepa\'{n}ska 1/5, \mbox{31-011} Cracow, Poland} \\[2ex]
		 Tomasz Miller\thanks {Corresponding Author: tomasz.miller@uj.edu.pl} \\ {\small Copernicus Center for Interdisciplinary Studies,} \\ {\small Jagiellonian University, Szczepa\'{n}ska 1/5, \mbox{31-011} Cracow, Poland}}

\maketitle

\begin{abstract}
     We generalize the differential space concept as a tool for developing differential geometry, and enrich this geometry with infinitesimals that allow us to penetrate into the superfine structure of space. This is achieved by Yoneda embedding a ring of smooth functions into the category of loci. This permits us to define a category of functorial differential spaces. By suitably choosing various algebras as ``stages'' in this category, one obtains various classes of differential spaces, both known from the literature and many so far unknown. In particular, if one chooses a Weil algebra, infinitesimals are produced. We study the case with the Weil algebra $\R \oplus \R[\epsilon^k]$ which allows us to fully develop the corresponding differential geometry with infinitesimals. To test the behavior of infinitesimals, we construct a simplified Robertson--Walker--Friedman--Lema{\^i}tre cosmological model.
\end{abstract}


\section{Introduction}

\label{intro}
The present paper has two objectives: first, to generalize and then to explore the concept of differential spaces as a tool for developing differential geometry of space-time; second, to enrich this geometry with infinitesimals that would allow us to penetrate into the superfine structure of space-time which is inaccessible to the standard geometric methods. This could prove interesting when dealing with the classical singularity problem, when the space shrinks to infinitesimal dimensions. These two objectives are not independent. Differential spaces encode the structure of space in a ring of functions rather than in atlases and maps, and rings of functions are a natural place for introducing infinitesimals, as it is done in synthetic differential geometry.

One of the first to introduce differential spaces were Postnikov \cite{Postnikov1,Postnikov2}, Sikorski \cite{Sikorski1967,Sikorski1971}, Spallek \cite{Spallek1967,Spallek1969} and Aronszajn \cite{Aronszajn1967,Aronszajn1980} (but see also \cite{MacLane1970}). Later on, many works and several generalizations of this concept were proposed (for a review of literature till 1992 see \cite{BuchnerHeller}).

Our starting point is a Sikorski differential space \cite{Sikorski1967,Sikorski1971}. It is a ringed space $(M,C)$, where $C$ is a set of real valued functions on $M$ satisfying the following conditions
\begin{enumerate}
\item[(a)] 
$\textrm{sc} \, C=C$, where $\textrm{sc} \, C:=\{\omega(f_1,...,f_n) \, | \, f_1,...,f_n\in C, \omega \in C^\infty({\mathbb{R}}^n)\}$,
\item[(b)]
$C_M=C$, where $C_M$ is the set of all local $C$-functions in the weakest topology $\tau_C$ in which all functions from $C$ are continuous.
\end{enumerate}
The ring $C$ is called a differential structure on $M$. Functions belonging to $C$ are --- \textit{ex definitione} --- smooth functions on $M$. If only (a) is satisfied, $(M,C)$ is called a Sikorski predifferential space.

If $M$ is a smooth manifold and $\Cinf(M)$ the ring of smooth functions on it, we obtain the Sikorski differential space $(M, \Cinf({M}))$, which can eventually serve as a model for space-time (see, e.g. \cite{HelSas1994,HelSas1995}). Our strategy develops as follows.

With the help of the Yoneda embedding we embed the differential structure $\Cinf(M)$ into the category of presheaves over the category $\LL $ of loci. The latter is defined as the opposite category of (finitely generated) smooth rings (details below), to obtain
\begin{align*}
\bar{M} := Y(\Cinf(M)) = \LL (- , \Cinf(M)).
\end{align*}
$\bar{M}$ is obviously a functor from $\LL^\mathrm{op}$ to $\mathrm{Sets}$. $\bar{M}$, with its differential structure $\Cinf (\bar{M})$ (see section 3 below), will be called a \textit{functorial differential space}. At the stage $A \in \LL $ we have
\begin{align*}
\bar{M}(A) := Y(\Cinf (M))(A),
\end{align*}
and we obtain a ringed space $(\bar{M}(A), \Cinf(\bar{M}(A)))$, called a \textit{generalized differential space}. This construction unifies various versions of differential spaces: by substituting for $A$ various objects of $\LL $, we obtain various classes of differential spaces, both known from the literature, and many hitherto unknown.

We shall especially be interested in functorial differential spaces with the Weil algebra of the form $\mathcal{W}_k = \R [\epsilon^k], \; k \in \{0\}\cup \NN , \; \epsilon^{k+1} = 0$, as its stage. Nilpotents, introduced in this way, can naturally be interpreted as infinitesimals. 

Infinitesimals of this class\footnote{They should not be confounded with infinitesimals introduced in the framework of the so-called nonstandard analysis.} appeared in Synthetic Differential Geometry (SDG) which formulates differential geometry in smooth topoi, interpreted as generalized smooth spaces. A smooth topos is a category, the objects of which behave like spaces, and one of the objects is the ``line object'' $R$, equipped with a commutative algebra structure with the property that for infinitesimal objects $S \subset R$ all morphisms $S \to R$ are linear \cite{SmoothTopos}. In this conceptual framework, infinitesimals are usually introduced axiomatically \cite{Lavend,KockSDG,KockManif}. However, our approach closely follows that of Moerdijk and Reyes \cite{MoerReyes} who have chosen a more algebraic line of reasoning. In fact, our approach reduces to that of Moerdijk and Reyes for instance when applied to $\Cinf(\R^n)$ (possibly divided by an ideal).

In SDG, there exist various classes of infinitesimals; we have preferred to work with those generated by the Weil algebra $\mathcal{W}_k = \R [\epsilon^k]$ since this allows us to explicitly calculate all relevant geometric formulae. In doing so, everything looks as if we worked within the usual world of sets, but if we tried to formally construct a model for our calculations, we would have to switch from the category of sets with its underlying classical logic to some ``smooth topos'' (which, in fact, we have done, since our $\bar{M}$ is  a full subcategory of the topos category $\mathrm{Sets}^{\LL^\mathrm{op}}$). As it is well known, the internal logic of topoi is the intuitionistic logic. We stick to this logic also by the fact that all our reasonings are strictly constructivist.

The program, outlined above, is implemented as follows. In section 2, we define the category $\Cinf $ of smooth rings, embed it in the category of loci, and define functorial differential spaces. In section 3, we consider finitely generated differential spaces as possible models of space-time, and develop differential geometry enriched with infinitesimals. In section 4, we construct a simple cosmological model in which, during all of its macroscopic evolution, infinitesimals remain latent, and become effective only in the close vicinity of the initial singularity. This model is intended only as a preliminary test-model for the role infinitesimals can play in the structure of the universe. Some concluding remarks are collected in section 5.

\section{Generalized differential spaces by Yoneda embedding}
\label{sec:1}
In this section, we prepare an environment in which functorial differential spaces with infinitesimals can be introduced.

\begin{definition}
A unital commutative $\R$-algebra $A$ is a $C^\infty $-ring if it is equipped with a smooth functional calculus, i.e., for any $n \in \NN ,\, \omega \in \Cinf(\R^n)$ and $a_1, \ldots , a_n \in A$, the element $\omega (a_1, \ldots , a_n)$ is defined and the following conditions are satisfied 
\begin{enumerate}
\item for $\phi, \psi \in C^\infty(\R^2),\quad \phi(x_1,x_2)= x_1\cdot x_2, \quad \psi(x_1,x_2)= x_1 + x_2$, 
\begin{align*}
\phi(a,b) = a\cdot b,\quad \psi(a,b) = a+b,
\end{align*}
\item for $\pi_i: \R^n \rightarrow \R, \ n \in\NN, \ \pi_i(x_1,\ldots,x_n)=x_i, \ i=1,2,\ldots,n, \ a_1,\ldots,a_n \in A$,
\begin{align*}
\pi_i(a_1,\ldots,a_n) = a_i,
\end{align*}
\item for the constant functions $1 \in C^\infty(\R^n), \ a_1,\ldots,a_n \in A$,
\begin{align*}
1(a_1,\ldots,a_n)= 1_A,
\end{align*}
\item for $\theta \in C^\infty(\R^m), \ \omega_1,\ldots,\omega_m \in C^\infty(\R^n), \ a_1,\ldots,a_n \in A, \ n,m \in \NN$,
\begin{align*}
(\theta \circ (\omega_1,\ldots,\omega_m))(a_1,\ldots,a_n)= \theta(\omega_1(a_1,\ldots,a_n),\ldots,\omega_m(a_1,\ldots,a_n)).
\end{align*}
\end{enumerate}
\end{definition}

Let $A,B$ be $C^\infty$-rings. A homomorphism $f: A \rightarrow B$ of $\mathbb{R}$-algebras is called $C^\infty$-morphism if, for any $\omega \in C^\infty(\mathbb{R}^n)$, $n \in\mathbb{N}$, $a_1,\ldots,a_n \in A$, the following equality is satisfied
\begin{align*}
f(\omega( a_1,\ldots,a_n )) = \omega(f(a_1),\ldots,f(a_n)).
\end{align*}

$C^\infty$-rings as objects with $C^\infty$-morphisms as morphisms form a category which will be denoted by $C^\infty $.

Of course, $\R $ is a $\Cinf $-ring with the operation: for any $\omega \in \Cinf (\R^n), \, x_1, \ldots , x_n \in \R $, the element $\omega(x_1, \ldots , x_n)$ is the value of $\omega $ for arguments $x_1, \ldots , x_n$.

\begin{definition}
$C^\infty$-ring $A$ is finitely (smoothly) generated by the set $\{g_1,...,g_n\}$ of elements $g_1,...,g_n \in A$ if every element $a\in A$ can be presented in the following way
\begin{align*}
a=\omega(g_1,...,g_n),
\end{align*}
for some $\omega \in C^\infty(\mathbb{R}^n)$. 
The set $\{g_1,...,g_n\}$ is called the set of generators of the $C^\infty$-ring $A$.
\end{definition}

Finitely generated $C^\infty$-rings (with suitable morphisms) form a subcategory of the category $C^\infty $. This subcategory will be denoted by $C^{\infty}_\mathrm{fg}$.

Algebras of the form $C^\infty(\R^n)/J$, where $J$ is an ideal of the underlying algebra, are good examples of finitely generated $C^\infty$-rings \cite{MoerReyes}.

In the following we shall consider the category, called the category of (smooth) loci, denoted by $\LL $, which is defined to be the dual category to the category of finitely generated $C^\infty $-rings, $\LL=(C^{\infty}_\mathrm{fg})^\mathrm{op}$. Although the objects of $\LL $ are the same as those of $(C^{\infty}_\mathrm{fg})$ (the arrows are reversed), we shall distinguish them by writing $\ell A \in \LL$ for $A \in C^{\infty}_\mathrm{fg}$.

Let us also notice that any $C^\infty$-ring $A$ can be recovered from the corresponding $\ell A$, since $A \cong \LL(\ell A, \ell C^\infty(\R ))$ (the operation of ``deleting'' $\ell $) \cite[p. 58]{MoerReyes}. 

Let now $A \in \Cinf $ (from now on when writing $\Cinf $ we always mean $C^\infty_\mathrm{fg}$, unless otherwise stated explicitly), and let us consider the Yoneda embedding
\begin{align*}
Y(\ell A) = \LL (-, \ell A).
\end{align*}
If also $B\in C^\infty $, we can write
\begin{align*}
Y(\ell A)(\ell B) = \LL (\ell B, \ell A) = \Hom_{\Cinf}(A,B).
\end{align*}
$\ell B$ is referred to as a stage from which $\ell A$ is regarded.

We now have all the necessary ingredients to define a generalized differential space as $(\Hom_{\Cinf }(A,B), \bar{A})$ where the differential structure  $\bar{A}$ of $\Hom_{\Cinf }(A,B)$ is defined in the following way. Let $a \in A$ and $\rho \in \Hom_{\Cinf }(A,B)$. We define
\begin{align*}
\bar{a}: \Hom_{\Cinf }(A,B) \to B
\end{align*}
by
\begin{align*}
\bar{a}(\rho ) = \rho (a) \in B,
\end{align*}
and finally
\begin{align*}
\bar{A} = \{\bar{a} \, | \, a \in A \}
\end{align*}
which is obviously a $\Cinf $-ring. We also have the operation: if $\omega \in \Cinf (\R^n)$ then we define 
\begin{align*}
\omega (\bar{a}_1, \ldots , \bar{a}_n) = \overline{\omega (a_1, \ldots , a_n)},
\end{align*}
for $a_1,\ldots , a_n \in A$.

\section{Geometry of differential spaces with infinitesimals}
\label{sec:2}
Let $(M, \Cinf (M))$ be a finitely generated differential space. We apply to $\Cinf (M) $ the Yoneda embedding $\bar{M} := Y(\ell \Cinf(M))$. Of course, $\bar{M} $ is a functor $\bar{M} : \LL^\mathrm{op} \to \mathrm{Sets} $. At a stage $\ell A \in \LL $ we have
\begin{align*}
\bar{M} (\ell A) = \Hom_{\Cinf }(\Cinf (M), A).
\end{align*}
Here, for any $f \in \Cinf (M)$, we define $\bar{f} := \{\bar{f}^A\}_{A \in \LL }$, $\bar{f}^A: \bar{M}(\ell A) \to A$ by $\bar{f}^A(\rho ) = \rho (f)$, $\rho \in \Hom_{\Cinf }(\Cinf (M), A)$, and
\begin{align*}
\Cinf (\bar{M}(\ell A)) = \{\bar{f}^A \, | \, f \in \Cinf (M) \}.
\end{align*}
$\Cinf (\bar{M}(\ell A))$ is a $\Cinf $-ring with the operation
\begin{align*}
\omega (\bar{f}_1^A, \ldots , \bar{f}_n^A) = \overline{\omega (f_1, \ldots, f_n)}^A,
\end{align*}
for $\omega \in \Cinf (\R^n)$.

We define $\Cinf(\bar{M}) = \{\bar{f} \, | \, f \in \Cinf(M)\}$. Let us notice that $\Cinf$-ring structure of $\Cinf(\bar{M})$ is defined ``pointwise'' in the following way: $\bar{f} + \bar{g} = \overline{f + g}$, $\bar{f} \cdot \bar{g} = \overline{f \cdot g}$, for $f, g \in \Cinf (M)$ and $\omega (\bar{f}_1, \cdots , \bar{f}_n) = \overline {\omega (f_1, \cdots , f_n)}$ for $\omega \in \Cinf (\R^n), \, f_1, \ldots , f_n \in \Cinf (M)$.

\begin{definition}
The pair $(\bar{M}, \Cinf(\bar{M}))$ is called a functorial differential space. At a stage $\ell A \in \LL$ it yields the generalized differential space $(\bar{M} (\ell A), \Cinf (\bar{M}(\ell A)))$.
\end{definition}

We define the category of functorial differential spaces in the following way. Its objects are functorial differential spaces $(\bar{M}, \Cinf (\bar{M}))$ and its morphisms
\begin{align*}
(\bar{M}, \Cinf (\bar{M})) \to (\bar{N}, \Cinf (\bar{N}))
\end{align*}
are the pairs of arrows
\begin{align*}
\bar{G}: \bar{M} \to \bar{N},
\end{align*}
for $G: M \to N$, and
\begin{align*}
\bar{G}^*: \Cinf (\bar{N}) \to \Cinf (\bar{M}),
\end{align*}
where $\bar{G}$ is defined as follows. At a stage $\ell A \in \LL $, the set $\bar{M} (\ell A)$ consists of arrows $\rho: \Cinf (M) \to A$, and $\bar{N} (\ell A)$ consists of arrows $\sigma: \Cinf (N) \to A$. We thus have the following commutative diagram

\begin{center}
\begin{tikzcd}
\Cinf(M) \arrow[leftarrow]{r}{G^\ast} \arrow[rightarrow]{d}[swap]{\rho} & \Cinf(N) \arrow[rightarrow]{ld}{\sigma}
\\
A & 
\end{tikzcd}
\end{center}
from which $\bar{G}^A(\rho ) = \rho \circ G^* = \sigma $ follows. 

The concept of functorial differential space is indeed very general. Having the functor
\begin{align*}
\bar{M}(\ell A) = \Hom_{\Cinf }(\Cinf(M), A),
\end{align*}
by suitably choosing  $\ell A$, we can recover various constructions known from the literature.

Let, for instance, $\ell A = \ell \Cinf (\R )$. We obtain
\begin{align*}
\bar{M} (\ell \Cinf (\R )) = \Hom_{\Cinf }(\Cinf (M), \Cinf (\R )) = \{ \phi^* \, | \, \phi : \R \to M \} \cong M^{\R }
\end{align*}
which is the family of smooth curves. In this way, we can recover the triple $(M, M^{\R }, \Cinf (M))$ which corresponds to Fr\"olicher spaces \cite{Frolicher}.

More generally, if $\ell A = \ell \Cinf (P)$, where $P$ is a smooth manifold, we obtain
\begin{align*}
\bar{M} (\Cinf (P)) = \Hom_{\Cinf } (\Cinf (M), \Cinf (P)) = \{\phi^* \, | \, \phi: P \to M \} \cong M^P,
\end{align*}
and we have a space, the points of which are parametrized by $P$ \cite{Navarro}.

Finally, let $M$ be a manifold and $\ast$ a point in $M$. Substituting $\ell A = \ell \Cinf (\{*\}) \cong \ell \R$ we obtain
\begin{align*}
\bar{M} (\ell \R) = \textrm{Spec}_r \Cinf (M) \cong M
\end{align*}
(subscript $r$ denotes ``real spectrum''), which gives Nestruev's approach \cite{Jet}.
 
Now, our aim is to construct the geometry of the functorial differential space of the functor $\bar{M} $ with the Weil algebra stage of the form $\mathcal{W}_k = \R [\epsilon^k ], \, k \in \NN $. To this end we consider, following Moerdijk and Reyes \cite{MoerReyes}, the quotient ring $R := \mathbb{R}[x]/(x^{k+1})$ with nilpotents of order $k$. Every element of $R$ can be presented in the following way
\begin{align*}
a = a_0 + d = a_0+ a_1\epsilon+a_2\epsilon^2+...+a_k\epsilon^k
\end{align*}
where $\epsilon$ is the equivalence class of $x$, $\epsilon^{k+1}=0$, $a_0,...,a_k\in \R$ and $d=a_1\epsilon+a_2\epsilon^2+...+a_k\epsilon^k$. The element $a_0$ is called the real part of $a$ and $d$ is called the nilpotent part of $a$. Of course $d^{k+1}=0$.
In fact $R$ is the direct sum $R = \mathbb{R} \oplus D_k$ of the linear space of the real numbers $\mathbb{R}$ and the linear space $D_k$ of nilpotents of order $k, k\in \mathbb{N}$.

In what follows, the object of our study will be the functorial differential space
\begin{align*}
(\bar{M}(k), \Cinf (\bar{M}(k))),\;\; k\in \NN ,
\end{align*}
abbreviated to $\bar{M}$, where
\begin{align*}
\bar{M}(k) = \Hom_{\Cinf }(\Cinf (M), \R [\epsilon^k]).
\end{align*}
In agreement with the general case considered above, for $f \in \Cinf (M)$ we have $\bar{f}(\rho ) = \rho  (f)$ with $\rho: \Cinf (M) \to \R [\epsilon ^k]$, and $\Cinf (\bar{M})$ is a $\Cinf $-ring with suitable operations.

\begin{proposition}
\label{isoring}
The mapping $J: C^\infty(M) \rightarrow C^\infty(\bar{M})$ given by
\begin{align*}
J(f)= \bar{f} \quad \mathrm{for} \quad f\in C^\infty(M),
\end{align*}
is an isomorphism of $C^\infty$-rings. 
\end{proposition}
{\em Proof.} Let $f,g \in C^\infty(M)$. It is easy to see that
\begin{align*}
\bar{f}=\bar{g} \ \Rightarrow \ f=g.
\end{align*}
Indeed, $\bar{f}(\rho)=f(p)+v_p(f)$ and $\bar{g}(\rho)=g(p)+v_p(g)$ for any $\rho \in \Hom_{\Cinf }(\Cinf (M), \R [\epsilon^k])$ with $f(p), g(p) \in \R$, $v_p(f), v_p(g) \in D_k$ and $p\in M$. Here, we have used the well-known fact that the only real-valued $\Cinf$-morphisms going from $\Cinf(M)$ are the evaluations \cite{Jet}. Thus $f=g$, and the mapping $J$ is a bijection satisfying $J(\omega(f_1,\ldots,f_n))= \omega(J(f_1),\ldots,J(f_n))$, for any $\omega \in C^\infty({\R}^n), \  f_1,\ldots, f_n\in C^\infty(M) $. Therefore, $J$ is an isomorphism of $C^\infty$-rings. \rule{5pt}{5pt}

\begin{corollary}
The $C^\infty(M)$-module of derivations $\Der(C^\infty(M))$ is isomorphic to the $C^\infty(\bar{M})$-module of derivations $\Der(C^\infty(\bar{M}))$.
\end{corollary}
{\em Proof.} For any derivation $X\in \Der(C^\infty(M)) $ we define $\bar{X}:C^\infty(\bar{M}) \rightarrow C^\infty(\bar{M})$ by
\begin{align*}
\bar{X}(\bar{f})=\overline{X(f)} \quad \mathrm{for} \quad f\in C^\infty(M).
\end{align*}
It is easy to see that $\bar{X} \in \Der(C^\infty(\bar{M}))$. Indeed, by Proposition \ref{isoring} we have $\bar{X}(J(f))=J(X(f))$, and $\bar{X}$ is a derivation as the composition of the isomorphism $J$ and the derivation $X$. 

It is easy to see the implication
\begin{align*}
\bar{X}=\bar{Y} \Rightarrow X=Y \quad \mathrm{for \ any} \ X,Y \in \Der(C^\infty(M)).
\end{align*}
Indeed, for any $f\in C^\infty(M)$ we have 
\begin{align*}
\bar{X}(\bar{f})=\bar{Y}(\bar{f}) \ \Rightarrow \ \overline{Xf} = \overline{Yf} \ \Rightarrow \ Xf = Yf
\end{align*}
and hence $X = Y$.

Therefore, the mapping $I: \Der(C^\infty(M)) \rightarrow \Der(C^\infty(\bar{M}))$, given by 
\begin{align*}
I(X)= \bar{X}  \quad \mathrm{for} \ X\in \Der(C^\infty(M)),
\end{align*}
is an isomorphism of modules. \rule{5pt}{5pt}
\\

This isomorphism allows us to construct differential geometry on spaces with infinitesimals.

\begin{definition}
For any linear connection $\nabla: \Der(C^\infty(M))\times\Der(C^\infty(M)) \rightarrow \Der(C^\infty(M))$, we define the linear connection $\bar{\nabla}: \Der(C^\infty(\bar{M}))\times \Der(C^\infty(\bar{M})) \rightarrow \Der(C^\infty(\bar{M}))$ by
\begin{align*}
\bar{\nabla}_{\bar{X}}\bar{Y}=\overline{\nabla_X Y} \quad \mathrm{for} \ X,Y \in \Der(C^\infty(M)).
\end{align*}
\end{definition}

In a similar manner, one can ``extend'' the usual definion of any tensor to a tensor on the space with infinitesimals. For example,
for any tensor $A: \Der(C^\infty(M))\times \cdots \times\Der(C^\infty(M)) \rightarrow \Der(C^\infty(M))$ of the type $(1,n)$, we define the tensor $\bar{A}:\Der(C^\infty(\bar{M}))\times \cdots \times \Der(C^\infty(\bar{M})) \rightarrow \Der(C^\infty(\bar{M}))$ by
\begin{align*}
\bar{A}(\bar{X_1}, \ldots ,\bar{X_n})=\overline{A(X_1, \ldots ,X_n)} \quad \mathrm{for} \ X_1, \ldots ,X_n\in \Der(C^\infty(M)).
\end{align*}

One can easily see that the following equalities are satisfied
\begin{align*}
\bar{X}=J\circ X, \quad \bar{\nabla}_{\bar{X}}\bar{Y}=J\circ \nabla_X Y, \quad \bar{A}(\bar{X}_1, \ldots,\bar{X}_n)=J\circ A(X_1, \ldots,X_n), \quad \bar{f}=J\circ f.
\end{align*}
where $J$ is the isomorphism of Proposition \ref{isoring}.

There is a one-to-one correspondence between geometric structures on $(M,C^\infty(M))$ and geometric structures on $(\bar{M},C^\infty(\bar{M}))$, in the sense that differential geometry on $(M,C^\infty(M))$ can be lifted to $(\bar{M},C^\infty(\bar{M}))$ and, conversely, differential geometry on $(\bar{M},C^\infty(\bar{M}))$ can be projected to $(M,C^\infty(M))$.

If $g: \Der(C^\infty(M)) \times \Der(C^\infty(M)) \rightarrow C^\infty(M)$ is a semi-Riemannian metric on $M$, we can consider the lift $\bar{g}: \Der(C^\infty(\bar{M})) \times \Der(C^\infty(\bar{M})) \rightarrow C^\infty(\bar{M})$ on $\bar{M}$. And if $\nabla $ is the Levi-Civita connection of $g$, then $\bar{\nabla} $ is the Levi-Civita connection of $\bar{g}$. The connection $\bar{\nabla }$ has torsion $\bar{\mathcal{T}}$ and curvature $\bar{R}$.

Let us now suppose that $(M,g)$ is a space-time on which the Einstein equations are defined
\begin{align*}
\Ric - \frac{1}{2} {\cal{R}}g + \Lambda g= 8 \pi T,
\end{align*}
where $\Lambda$ is the cosmological constant, $\Ric$ is the Ricci curvature (a symmetric (0,2) tensor), ${\cal{R}} \in C^\infty(M)$ is the scalar curvature and $T$ a suitably defined energy-momentum tensor. By using the above presented machinery these equations can be lifted to the space-time $(\bar{M}, \bar{g})$ with infinitesimals to obtain
\begin{align*}
\overline{\Ric}  - \frac{1}{2} \bar{{\cal{R}}} \bar{g}  + \Lambda \bar{g}= 8 \pi \bar{T} .
\end{align*}

As we can see, nothing essentially new has been obtained by taking into account the smallest (infinitesimal) structure of space-time. This had to be expected. The standard differential geometric tools are taylored to deal with the macroscopic structure of space-time. To penetrate into its fine-grained structure, we must go beyond the ``isomorphism consisting of overlining macroscopic structures''.

\section{RWFL evolution of the universe with nilpotents}
\label{sec:3}
In this section we consider the Robertson--Walker--Friedman--Lema{\^i}tre (RWFL) evolution of the universe, in order to see how the presence of infinitesimals influences the structure of the initial singularity.

Let $M = \itDelta \times_{\mathcal{S}} \itSigma$ be the usual RWFL space-time (see, e.g., \cite{ONeill}). Here $\itDelta = [0, \infty ) \subset \R $ is the ``time axis'', $\itSigma$ is a three dimensional Riemannian manifold of constant curvature with the curvature parameter $\kappa \in \{0, \pm 1\}$, and $\mathcal{S}: \itDelta \to \R$ is a scaling function. There are two projections: $\pi_1: \itDelta \times \itSigma \to \itDelta$ and $\pi_2: \itDelta \times \itSigma \to \itSigma$. $M$ carries the metric
\begin{align*}
g = -\pi_1^* (dt)^2 + \mathcal{S}^2 \pi_2^*g.
\end{align*}

The model has the initial singularity. It occurs at $t_* = 0$ when $\mathcal{S} \to 0$ and $\mathcal{S}' \to \infty $ as $t \to 0^+$.

Everything so far is like in the standard description of the RWFL model. Now, we want to introduce infinitesimals into the model. In accordance with our previous notation,
\begin{align*}
\bar{M} = \{\bar{M}(k)\}_{k\in \NN},
\end{align*}
where $\bar{M}(k) = \Hom_{\Cinf }(\Cinf (M), \R [\epsilon ^k])$. If $ \bar{M}(k) \ni \rho : \Cinf (M) \to \R [\epsilon^k]$, where $\R [\epsilon^k]$ is a Weil algebra of the form $\R \oplus D_k$, then $\rho = \ev_p + v_p$, $p \in M$, where $\ev_p: \Cinf (M) \to \R $ is the evaluation at $p$, and $v_p: \Cinf (M) \to D_k$  an ``infinitesimal tangent vector'' of order $k$ at $p$.\

We now define the differential structure on $\bar{M}(k)$ to be
\begin{align*}
\Cinf (\bar{M}(k)) = \{\bar{f}^{(k)} \, | \, f \in \Cinf(M) \},
\end{align*}
where $\bar{f}^{(k)} := \bar{f}^{\R [\epsilon^k]}$, with $\bar{f}^{(k)}(\rho ) = \rho (f) = f(p) + v_p(f)$ for $\rho \in \bar{M}(k)$. For $f \in \Cinf (M)$, we have here defined $\bar{f} = \{\bar{f}^{(k)}\}_{k \in \NN }$ as a smooth function on $\bar{M}, \; \bar{f}(\rho ) := \bar{f}^{(k)}(\rho)$.

We must prepare some tools to deal with infinitesimals. Let $\rho,\sigma \in \bar{M}(k)$. We define the ``$k$-th order neighbouring relation'' \cite[chapter 2]{KockManif}.
\begin{align*}
\rho \sim_k \sigma \ \Leftrightarrow \ \sigma - \rho \in D_k,
\end{align*}
where 
\begin{align*}
D_k = \{d \in R \, | \, d^{k+1} = 0\}
\end{align*}
This relation is reflexive and symmetric, but instead of transitivity we have
\begin{align*}
(\rho \sim_k \sigma \wedge \sigma \sim_l \tau) \ \Rightarrow \ (\rho \sim_{k+l} \tau).
\end{align*}

These properties, being similar to those of a distance function, suggest to consider a ``quasi-distance'' function satisfying the inequality \cite[p. 83]{KockManif}
\begin{align*}
\mathrm{qdist}(\rho , \sigma) \leq k \;\; \mathrm{if} \;\; \rho\sim_k \sigma.
\end{align*}
Since $D_k(n) \subseteq D_l(n)$, provided $k \leq l$, the function ``qdist'' determines a ``size'' of an infinitesimal object $D_k$.

With the help of qdist-function we define another useful concept. For $\rho \in \bar{M}$, the $k$-monad around $\rho $ is defined to be
\begin{align*}
\cM_k(\rho ) := \{\sigma \in \bar{M}(k) \, | \, \rho\sim_k \sigma\}.
\end{align*}
For $k=1$, we write $\cM(\rho)$. Obviously, $\sigma \in \cM_k(\rho)$ if and only if $\rho \in \cM_k(\sigma)$. 

We agree to define $D_{\infty } = \bigcup_{k=1}^{\infty }D_k$, i.e., $\rho \sim _{\infty } \sigma \Leftrightarrow \exists_{k \in \NN } \, \sigma - \rho \in D_k $. We also write
\begin{align*}
\mathcal{M}_{\infty }(\rho) = \{\sigma \in \bar{M}(k) \, | \, \rho \sim_{\infty } \sigma \}.
\end{align*}

Let us also notice that, for $\rho = \ev_p + v_p$, $p \in M$, we have $\mathcal{M}_k(\rho) \cong D_k$ and $\mathcal{M}_{\infty }(\rho) \cong D_{\infty }$.

Let us now consider the evolution of the RWFL cosmological model backwards in time starting from the present epoch. We assume that its space-time is smooth (in the usual sense) which means that we have a bundle of monads $\{\mathcal{M}_{\infty }(\ev_p)\}_{p \in M}$ over the space-time manifold $M$ (in $M$, there exist infinitesimals for any $k$). Everything goes according to the standard scenario: space shrinks, density and temperature grow. Throughout the evolution Proposition \ref{isoring} remains valid, i.e., rings $\Cinf(M)$ and $\Cinf(\bar{M}(k))$ are isomorphic and, consequently, the inclusion of infinitesimals does not import macroscopically visible effects. Something like that should be expected since there is no reason why infinitesimals would have any role to play in a macroscopic evolution.

However, when contraction goes sufficiently far, one should face the existence of the initial singularity. As is well known, this means that there must exist at least one incomplete timelike or null geodesic (an incomplete causal curve) that cannot be continued in any extension of space-time.  In the case of the RWFL model, the scalar $R_{ijkl}R^{ijkl}$ constructed from the curvature tensor, becomes unbounded on approaching the singularity (in RWFL models, the singularity is in fact a strong curvature singularity \cite{EllisSchmidt}). As the curvature increases, the differentiability of space-time deteriorates and, finally, breaks down\footnote{There are strong reasons to claim that in the ``infinitesimal regime'' close to the singularity the curvature assumes only infinitesimal values \cite{HellerKrol}.}. This is one of the most intricate problems in the theory of classical singularities  (e.g., \cite{Clarke}).

Let us assume that everything goes on classically till the space is contracted to a single point $*$. When this point is reached, the fibres $\mathcal{M}_{\infty }(\ev_p), \; p \in M$, are reduced to the fibre $\mathcal{M}_k(\ev_\ast)$ where $k \neq \infty $. The ring of functions over $\{*\}$ is of course $\Cinf(\{*\}) \cong \R $ and, in agreement with our model, we have
\begin{align*}
\bar{M}(k) = \Hom_{\Cinf }(\R , \R[\epsilon^k]).
\end{align*}

If $k>0$, the evolution can proceed further, with $k$ becoming a smaller and smaller nonnegative integer (let us notice that this is a strong ``if''). 
We thus have the sequence of decreasing monads
\begin{align*}
\mathcal{M}_{k}(\ev_\ast) \supseteq \mathcal{M}_{k-1}(\ev_\ast) \supseteq \mathcal{M}_{k-2}(\ev_\ast) \supseteq \ldots
\end{align*}

This can be interpreted in the following way. When the point $*$ is reached, the usual time $t$ loses its meaning, and the ``duration'' starts to be measured by another ``quantised'' time parameter, namely, by the decreasing sequence of $k$'s. Finally, when $k= 0$, we have the monad $\mathcal{M}_0(\ev_\ast)$ for which
\begin{align*}
\rho \sim_0 \sigma \ \Leftrightarrow \ \sigma - \rho \in D_0 \ \Rightarrow \ \rho = \sigma,
\end{align*}
and everything reduces to a single point. 

Finally, let us notice that everything, described above, happens in the same category, i.e., the category of functorial differences spaces with infinitesimals.

\section{Concluding remarks}
\label{sec:concluding_remarks}
The first result of the present work is the formulation of a very general scheme --- the category of functorial differential spaces --- unifying many, known so far, conceptions of differential spaces, and  enabling to produce new ones. The scheme has a strong, both differential and algebraic, aspects. This is well resonant with the present tendency of geometrization and algebraization in contemporary theoretical physics.

The category of functorial differential spaces can indeed serve as a machine generating various spaces: it is enough to look at a certain differential structure from various stages. If such a stage is a suitable Weil algebra, infinitesimals inevitably appear. We have worked with the Weil algebra $\mathcal{W}_k = \R \oplus \R [\epsilon^k]$, but there are many other possibilites.

To successfully apply a mathematical theory to physics two things are required: an advanced mathematical theory and sufficiently elaborated calculatory techniques related to this theory. The former guarantees a theoretical insight into studied phenomena; the latter allows for translating this insight into the possibility of predicting some new effects. Synthetic differential geometry is an advanced mathematical theory \cite{Lavend,KockSDG,KockManif}, but has only a few applications in physics (see, e.g, \cite{Nashimura1,Nashimura2,Guts1,Guts2,Guts3,Guts4}). It seems that the reason for this is the lack of well elaborated methods of doing concrete calculations. And here comes the second main result of the present paper: working with the concrete Weil algebra $\mathcal{W}_k = \R \oplus \R [\epsilon^k]$ puts an abstract mathematical theory in the form ready for calculations.

It seems obvious that the existence of infinitesimals should be especially attractive in studying space-time in its microscale, and this should come into play when dealing with the classical\footnote{I.e., without taking into account quantum gravity effects.} singularity problem in cosmology and astrophysics. The RWFL world model with the initial singularity, constructed in this work, is only a toy model, but it shows how infinitesimals can do their work in regions in which the usual methods are ineffective.

Our Proposition \ref{isoring} implies --- rather an obvious conclusion! --- that at the macro-level the geometry of infinitesimals remains invisible. However, this does not mean that on the scale at which the microstructure of space-time and quanta are expected to interact with one another it cannot play an essential role. Our approach is in line with the ongoing program employing topos theory --- or, more generally, categorical methods --- to quantum physics (see, e.g., \cite{Isham2008,Landsman2009,Bohrif,Raptis1}), and the existence of infinitesimals belongs to the internal logic of many categories, topoi in particular (in \cite{Butter}, the role of infinitesimals in quantum mechanics and quantum gravity is explicitly discussed). 

Once infinitesimals have so naturally appeared in mathematics, it would be a great neglect not to use them to solve pressing problems in physics.

\section*{Acknowledgments}
We thank Ryszard Kostecki for a careful perusal of the manuscript and insightful comments.

\end{document}